\documentclass{jfm}
\usepackage{natbib}
\usepackage{graphicx}
\usepackage{subfigure}
\ifCUPmtlplainloaded \else
  \checkfont{eurm10}
  \iffontfound
    \IfFileExists{upmath.sty}
      {\typeout{^^JFound AMS Euler Roman fonts on the system,
                   using the 'upmath' package.^^J}%
       \usepackage{upmath}}
      {\typeout{^^JFound AMS Euler Roman fonts on the system, but you
                   dont seem to have the}%
       \typeout{'upmath' package installed. JFM.cls can take advantage
                 of these fonts,^^Jif you use 'upmath' package.^^J}%
      }
  \else
  \fi
\fi

\ifCUPmtlplainloaded \else
  \checkfont{msam10}
  \iffontfound
    \IfFileExists{amssymb.sty}
      {\typeout{^^JFound AMS Symbol fonts on the system, using the
                'amssymb' package.^^J}%
       \usepackage{amssymb}%

      }{}
  \fi
\fi

\ifCUPmtlplainloaded \else
  \IfFileExists{amsbsy.sty}
    {\typeout{^^JFound the 'amsbsy' package on the system, using it.^^J}%
     \usepackage{amsbsy}}
    {\providecommand\boldsymbol[1]{\mbox{\boldmath $##1$}}}
\fi

\newcommand\omv{\boldsymbol{\omega}}

\newsavebox{\astrutbox}
\sbox{\astrutbox}{\rule[-5pt]{0pt}{20pt}}

\title{Isotropic third-order statistics in turbulence with helicity:
the 2/15-law} 

\author[S. Kurien, M.A~Taylor, T. Matsumoto]{%
S\ls U\ls S\ls A\ls N\ns K\ls U\ls R\ls I\ls E\ls N$^1$,\ns% 
M\ls A\ls R\ls K\ns A.\ns T\ls A\ls Y\ls L\ls O\ls
R$^2$\thanks{Present address: Evolutionary Computing, Sandia National
  Laboratory, Albuquerque 87185, USA}
\and T\ls A\ls K\ls E\ls S\ls H\ls I\ns M\ls A\ls T\ls S\ls U\ls M\ls O\ls
T\ls O$^3$}

\affiliation{$^1$ Center for Nonlinear
Studies and Theoretical Division, Los Alamos National Laboratory,
Los Alamos, NM 87545, USA \\[\affilskip] 
$^2$ Computer and Computational Sciences Division, Los
Alamos National Laboratory, Los Alamos, NM 87545, USA \\[\affilskip]
$^3$ Department of Physics, Kyoto University, Kitashirakawa Oiwakecho
  Sakyo-ku, Kyoto 606-8502, Japan \\[\affilskip]}

\date{\today}

\begin{document}

\maketitle

\begin{abstract}
  The so-called 2/15-law for two-point, third-order velocity
  statistics in isotropic turbulence with helicity is computed for the
  first time from a direct numerical simulation of the Navier-Stokes
  equations in a $512^3$ periodic domain. This law is a statement of
  helicity conservation in the inertial range, analogous to the
  benchmark Kolmogorov 4/5-law for energy conservation in
  high-Reynolds number turbulence. The appropriately normalized
  parity-breaking statistics, when measured in an arbitrary direction
  in the flow, disagree with the theoretical value of 2/15 predicted
  for isotropic turbulence. They are highly anisotropic and variable
  and remain so over a long times. We employ a recently developed
  technique to average over many directions and so recover the
  statistically isotropic component of the flow. The angle-averaged
  statistics achieve the 2/15 factor to within about $7\%$
  instantaneously and about $5\%$ on average over time. The inertial-
  and viscous-range behavior of the helicity-dependent statistics and
  consequently the helicity flux, which appear in the 2/15-law, are
  shown to be more anisotropic and intermittent than the corresponding
  energy-dependent reflection-symmetric structure functions, and the
  energy flux, which appear in the 4/5-law. This suggests that the
  Kolmogorov assumption of local isotropy at high Reynolds numbers
  needs to be modified for the helicity-dependent statistics
  investigated here.

%\centering{\bf WORK IN PROGRESS, NOT FOR DISTRIBUTION}

\end{abstract}
%\pacs{}

\section{Introduction}
There are two invariants of the inviscid Navier-Stokes equations --
the total energy, defined by $E = \frac{1}{2}\int {\bf u}({\bf x})^2
d{\bf x}$, and the total helicity $H =\int {\bf u}({\bf x})\cdot
\omv({\bf x}) d{\bf x}$ where the vorticity $\omv({\bf x})=\nabla
\times {\bf u}({\bf x})$. Energy has been extensively studied especially in
statistical theories of turbulence as well as in experiments. Helicity, being
sign-indefinite has been more challenging to study
theoretically. Direct experimental measurements of helicity are also
difficult because of the need to measure local gradients, requiring
high resolution and careful probe design (see for example
\cite{KhoShaTsi01}). Nevertheless, since the discovery of helicity as
a conserved quantity by \cite{Moreau} and independently by
\cite{Moffat}, there have been several attempts to draw parallels with
the energy dynamics. The existence of a helicity cascade was proposed
by \cite{BFLLM73} and various possible inertial range scalings of the
energy and helicity spectra were discussed. The joint forward
(downscale) cascade of energy and helicity has been verified in direct
numerical simulations, most recently by \cite{CCE03}. A recent work by 
\cite{KurTayMat04}, showed that there is a relevant timescale for
helicity transfer in wavenumber space. The proper consideration of the
helicity flux timescale showed that helicity can modify the energy
dynamics, measureably slowing it down in the high wavenumbers.

We here present a study of the small-scale phenomenology of turbulence
with helicity in the manner of the \cite{K41b} investigation (K41)
of helicity-free turbulence.  Using the K\'arm\'an-Howarth
equation for the dynamics of the second-order two-point correlation
function (see \cite{KH38}), K41 gives the benchmark 4/5
energy law for homogeneous, isotropic, reflection-symmetric
(helicity-free) turbulence, assuming finite mean energy
dissipation $\varepsilon$ as $\nu \rightarrow 0$,
\begin{equation}
\langle (u_L({\bf x}+{\bf r}) - u_L({\bf x}))^3 \rangle =
-\frac{4}{5}\varepsilon r
\label{k45}
\end{equation}
for $\eta \ll r \ll L_0$, the so-called inertial range. See
\cite{Onsager} for a recently discovered derivation of a similar law
by Onsager. $\eta$ is the Kolmogorov dissipation scale, $L_0$ is the
typical large scale, $u_L({\bf x}) = {\bf u}({\bf x})\cdot {\hat {\bf
    r}}$ is the longitudinal component of ${\bf u}$ along ${\hat {\bf
    r}}$, the mean energy flux in the inertial range equals the mean
dissipation rate $\varepsilon = 2\nu \langle |\nabla {\bf
  u}|^2\rangle$, and $\langle \cdot \rangle$ denotes an ensemble
average of a high-Reynolds number decaying flow. It has been shown
empirically and proved that this is equivalent to a long-time average
in statistically steady high-Reynolds number turbulence
(\cite{Frisch95}). The 4/5-law is a statement of the conservation of
energy in the inertial range scales -- the third-order structure
function is an indirect measure of the flux of energy through scales
of size $r$. A key assumption of the K41 theory was `local isotropy'
or isotropy of the small scales $r \ll L_0$ at sufficiently high
Reynolds number. This assumption appears to hold according to high
Reynolds number experimental measurements of the 4/5-law even when the
data are acquired in only a single direction in the flow
(\cite{SreDhr98}).

Recently, a local version of the K41 statistical laws were proved in
\cite{DuchonRobert00} (see also \cite{Eyink03} for the case of the 4/5-law in
particular): Given $any$ local region $B$ of size $R$ of the flow,
for $r\ll R$, and in the limits $\nu \rightarrow 0$, next
$r\rightarrow 0$, and finally $\delta\rightarrow 0$,
\begin{eqnarray} 
\langle (\Delta u_L)^3 \rangle_{(\Omega,B)} &=& 
\lim_{\delta\rightarrow 0}\frac{1}{\delta}\int^{t+\delta}_t d\tau
\int \frac{d\Omega}{4\pi} 
 \int_B\frac{d{\bf x}}{R^3} [\Delta u_L({\bf r};{\bf x},\tau)]^3 
= -\frac{4}{5}\varepsilon_B r. \label{eyink45}
\end{eqnarray}
for almost every (Lebesgue) point $t$ in time, where $\Delta u_L({\bf
r}) = u_L({\bf x}+{\bf r}) - u_L({\bf x})$ and $\varepsilon_B$ is the
instantaneous mean energy dissipation rate over the local region
$B$. The angle integration $d \Omega$ integrates in ${\bf r}$ over the
sphere of radius $r$. For each point ${\bf x}$ the vector increment
${\bf r}$ is allowed to vary over all angles and the resulting
longitudinal moments are integrated. The integration over ${\bf x}$ is
over the flow subdomain $B$. This version of the K41 result does not
require isotropy, homogeneity, long-time or ensemble averages, or 
stationarity of the flow. This version of the 4/5-law has not yet been
rigorously verified empirically. It was shown by \cite{TayKurEyi03}
that at the very least, the 4/5-law does not seem to require isotropy,
long-time or ensemble averaging; it appears to be sufficient to 
average over many angles and over the domain at any instant of a
sufficiently `high' Reynolds number flow.

%averaging the third-order structure function over a large
%(finite) number of directions, that is, allowing ${\bf {\hat r}}$ to
%vary over angles for a fixed separation length $r$, gave results
%consistent with the 4/5-law at any given instant of the flow. 

The first attempt to study the symmetry and dynamics of the two-point
correlation function in flows with helicity was made by
\cite{Betchov61}. The simplest symmetry breaking of a statistically 
rotationally invariant flow is to break parity (mirror-symmetry) by the
introduction of helicity. A useful analogy which we borrow from
\cite{Betchov61} is of a well-mixed box of screws. This is
statistically invariant to rigid rotations, that is it is isotropic.
But under reflection in a mirror all the left-handed screws become
right-handed and vice-versa, that is the system is parity- or mirror-symmetry
breaking.  In particular, no combination of rigid rotations can
transform the box of screws into its reflected image. This is the type
of symmetry breaking we are considering here. Analogous to the K41
4/5-law, the so-called $2/15$-law for homogeneous, isotropic
turbulence with helicity was derived by \cite{Chkhetiani96} (see also
\cite{LPP97} and \cite{Kurien03}),
\begin{equation}
\langle \Delta u_L({\bf r}) (u_T({\bf x}+{\bf r}) \times u_T({\bf x}))
\rangle = \frac{2}{15} h r^2
\label{215law}
\end{equation}
where the transverse component of the velocity 
$u_T({\bf x}) = {\bf u}({\bf x}) - u_L({\bf x})$; 
the mean helicity flux which equals the mean helicity dissipation rate 
in steady state is 
$h=2\nu\langle (\partial_j u_i)(\partial_j \omega_i)\rangle$, 
where the vorticity $\omv = \nabla \times {\bf u}$. 
We shall use the notation
\begin{equation}
H_{LTT}({\bf r}) = \langle \Delta u_L({\bf x}) (u_T({\bf x}+{\bf r}) 
\times u_T({\bf  x})) \rangle 
\end{equation}
to denote the third-order helical statistics.  The quantity
$H_{LTT}({\bf r})$ is the simplest third-order velocity correlation
which can have a spatially isotropic component while at the same time
displaying a `handedness' due to the cross-product in its definition.
$H_{LTT}({\bf r})/r^2$ is a measure of the helicity flux through
scales of size $r$ in the inertial range which must balance the
helicity dissipation $h$ in the viscous range for statistically steady
turbulence. The 2/15-law assumes inertial-range behavior of helicity
in some range of scales $\eta \ll r \ll L_0$. A shell model
calculation by \cite{BifPieTos98} has demonstrated the likelihood of
the 2/15-law. However it has not been measured in experiments or,
until the present work, in direct numerical simulations of the
Navier-Stokes equations.
  
Theoretically, the 2/15-law has been shown to apply to the case of
high-Reynolds number decaying flows; the arguments for the forced case
have not yet been developed. In our initial investigation into the
2/15-law statistics for decaying flows with prescribed isotropic
helicity and energy spectra (see \cite{PolShtil89}), we were able to
observe some of the qualitative features of the flow but the Reynolds
numbers achievable for given our computing abilities was insufficient
to observe the 2/15-law. We therefore moved to perform forced
simulations to achieve higher Reynolds numbers and to use the
statistically steady state to compute the statistics.
  
In the next section describes the simulations and the calculation of
the statistical quantities of interest for the 2/15-law. We present a
comparison with the 4/5-law calculation of the same flow, highlighting
the differences between energy and helicity dynamics.  We show that in
the inertial range the helicity flux is more anisotropic and
intermittent than the energy flux; and that the smallest resolved
scales show recovery of isotropy for energy-dependent statistics but
show persistent anisotropy for helicity-dependent statistics over the
10 large-eddy turnover times for which simulation ran. We will
conclude with some final remarks on what our analysis means for future
work in the area of helicity dynamics and parity-breaking in turbulent
flows.

\section{Simulations and Results}
% unit-domain units
\begin{table}
\centering
\begin{tabular}{cccccccc}
$N$& $\nu$& $R_\lambda$ &$E$ & $\varepsilon$ & $H$
& $h$ &$k_{\rm max}\eta$  \\[3.5pt]
512 & $10^{-4}$& 270& 1.72 & 1.51 & -26.8 & 62.2 & 1.1 \\
\end{tabular}
\caption{Parameters of the numerical simulation. $\nu$ - viscosity;
  $R_\lambda$ - Taylor Reynolds number; 
  mean total energy $E = \frac{1}{2}\sum_k \tilde {\bf u}({\bf k})^2$; 
  $\varepsilon$ - mean energy dissipation rate;
  mean total helicity $H = \sum_k \tilde {\bf u}({\bf k}) \cdot \tilde 
  {\omv}({\bf k})$; $h$ - mean helicity dissipation rate; Kolmogorov
  dissipation scale $\eta = (\nu^3/ \varepsilon)^{3/4}$.}
\label{params}
\end{table}
\begin{figure}
  \centering
  \includegraphics[scale=0.35]{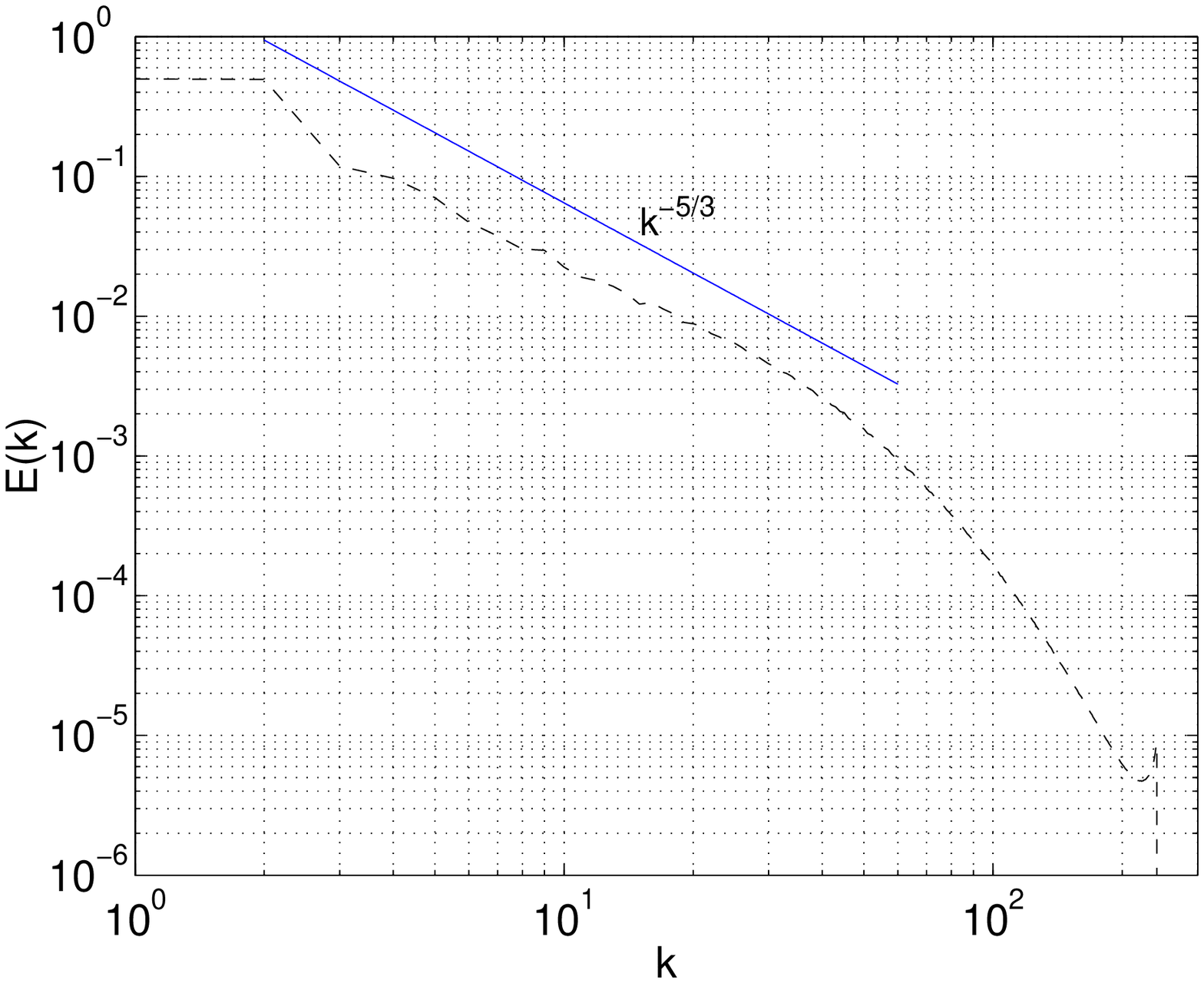}
  \includegraphics[scale=0.35]{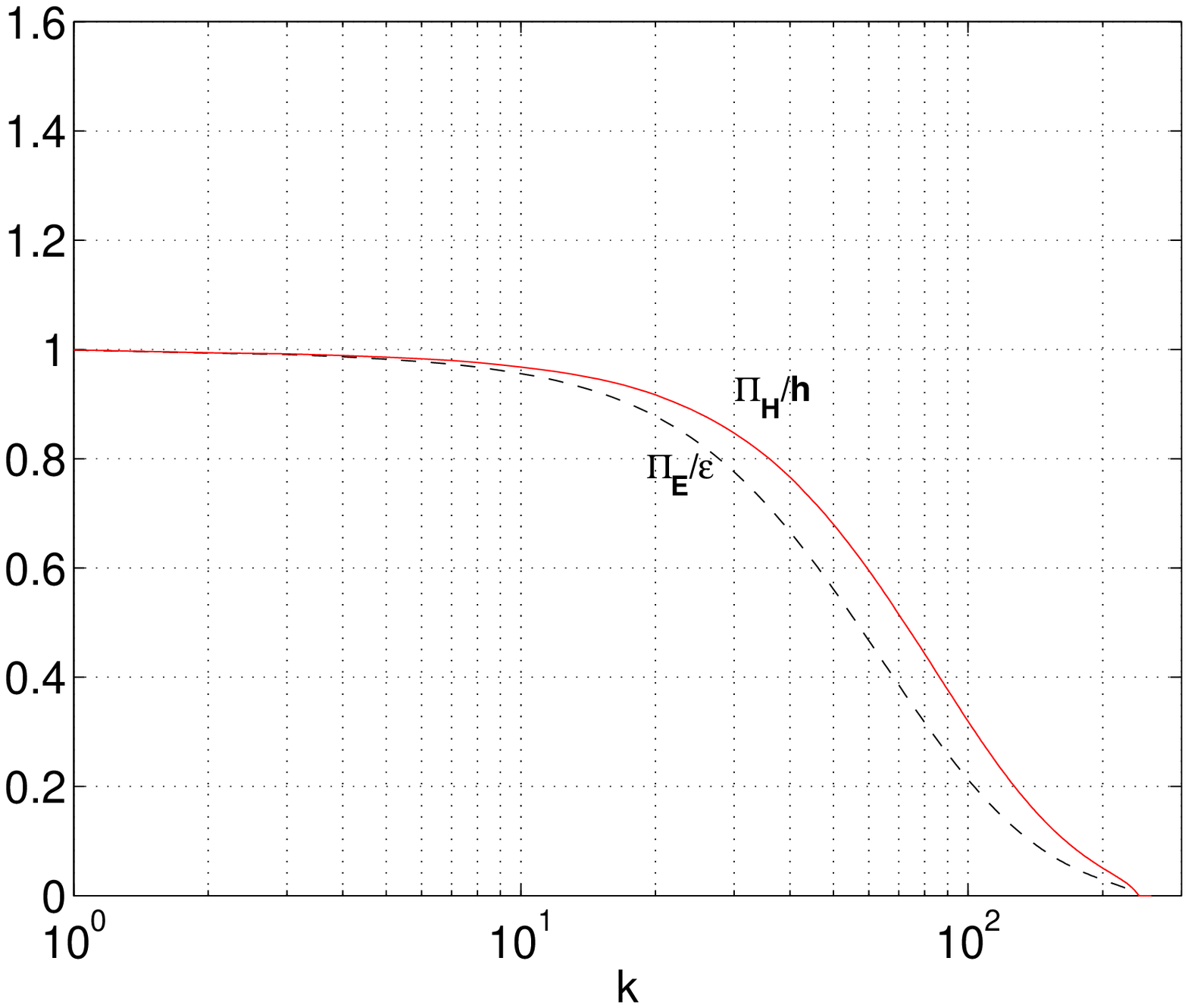}
\caption{Left panel: Dashed line -- Energy spectrum as a function of
  wavenumber. Solid line -- Line of slope -5/3 on the log-log scale.
  Right panel: Dashed line -- Flux of energy $\Pi_E$ normalized by
  mean dissipation rate of energy $\varepsilon$. Solid line -- Flux of
  helicity $\Pi_H$ normalized by mean dissipation rate of helicity
  $h$.\label{flow-params}}
\end{figure}
We performed a simulation of the forced Navier-Stokes equation in a
unit-periodic box with $512$ grid points to a side. In these units the
wavenumber $k$ is in integer multiples of 2$\pi$. The forcing scheme
was the deterministic forcing of \cite{TayKurEyi03}, modeled after the
deterministic forcing used in \cite{CCE03}.  This forcing simply
relaxes the Fourier coefficients in the first two wave numbers so that
the energy matches a prescibed target spectrum $F(k) = 0.5$ ($k=1,2$).
The forcing does not change the phases of the coefficients, which are
observed to change slowly in time. In addition, maximum helicity was
injected into the wavenumbers 1 and 2 using the scheme of
\cite{PolShtil89}. The calculation ran for 10 large-eddy turnover
times. The flow achieved steady state in about 1 large-eddy turnover
time. The statistics reported here have been calculated over a total
of 45 frames spanning the latter 9 eddy turnover times.  The same data
were reported in \cite{KurTayMat04}. Some additional parameters of the
simulation are given in Table~\ref{params}.  The left panel of
Fig.~\ref{flow-params} shows the mean energy spectrum with a line
indicating the Kolmogorov $k^{-5/3}$ scaling and (see
\cite{KurTayMat04} for the interpretation of the deviation from
$k^{-5/3}$ at the high end of the inertial range) The right panel of
Fig.~\ref{flow-params} shows helicity fluxes normalized by the mean
energy and helicity dissipation rates respectively. Note the close to
decade range of wavenumbers where $\varepsilon$ and $h$ match the
energy and helicity fluxes respectively very well.

\subsection{Third-order helical velocity statistics and the use of angle-averaging}
\begin{figure}
  \centering \subfigure[$\widetilde H_{LTT}({\bf r})$ in 73 different
  directions of the flow (dotted lines) from a single frame after the
  flow has reached statistically steady state. The thick solid line is
  the mean over all the
  directions.]{\label{k215_frame}\includegraphics[scale=0.35]{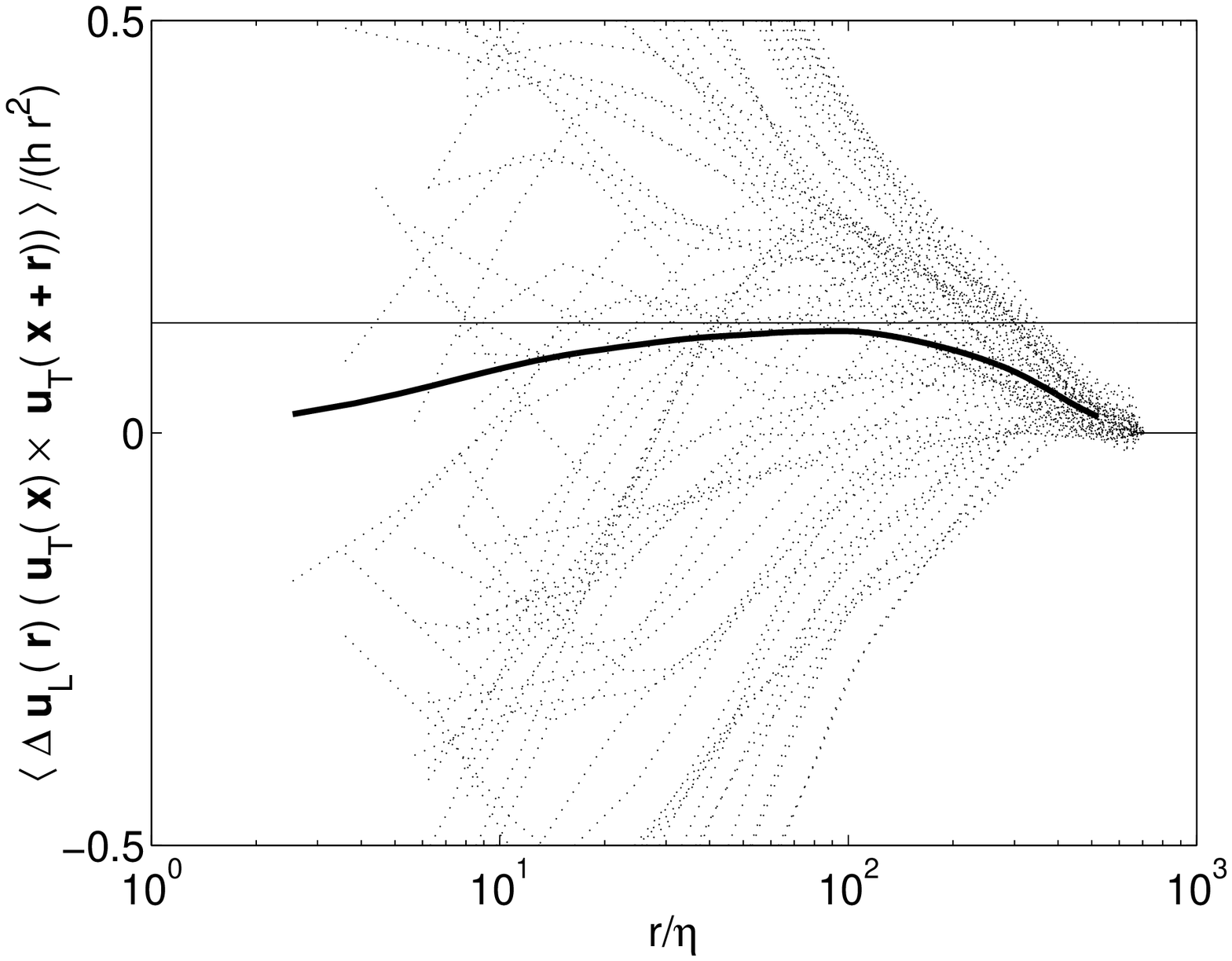}}
  \subfigure[The time-average of $\widetilde H_{LTT}({\bf r})$ in each
  of the 73 different directions (dotted lines) and of the
  angle-average (thick solid line).]  {\label{k215_timeave}
    \includegraphics[scale=0.35]{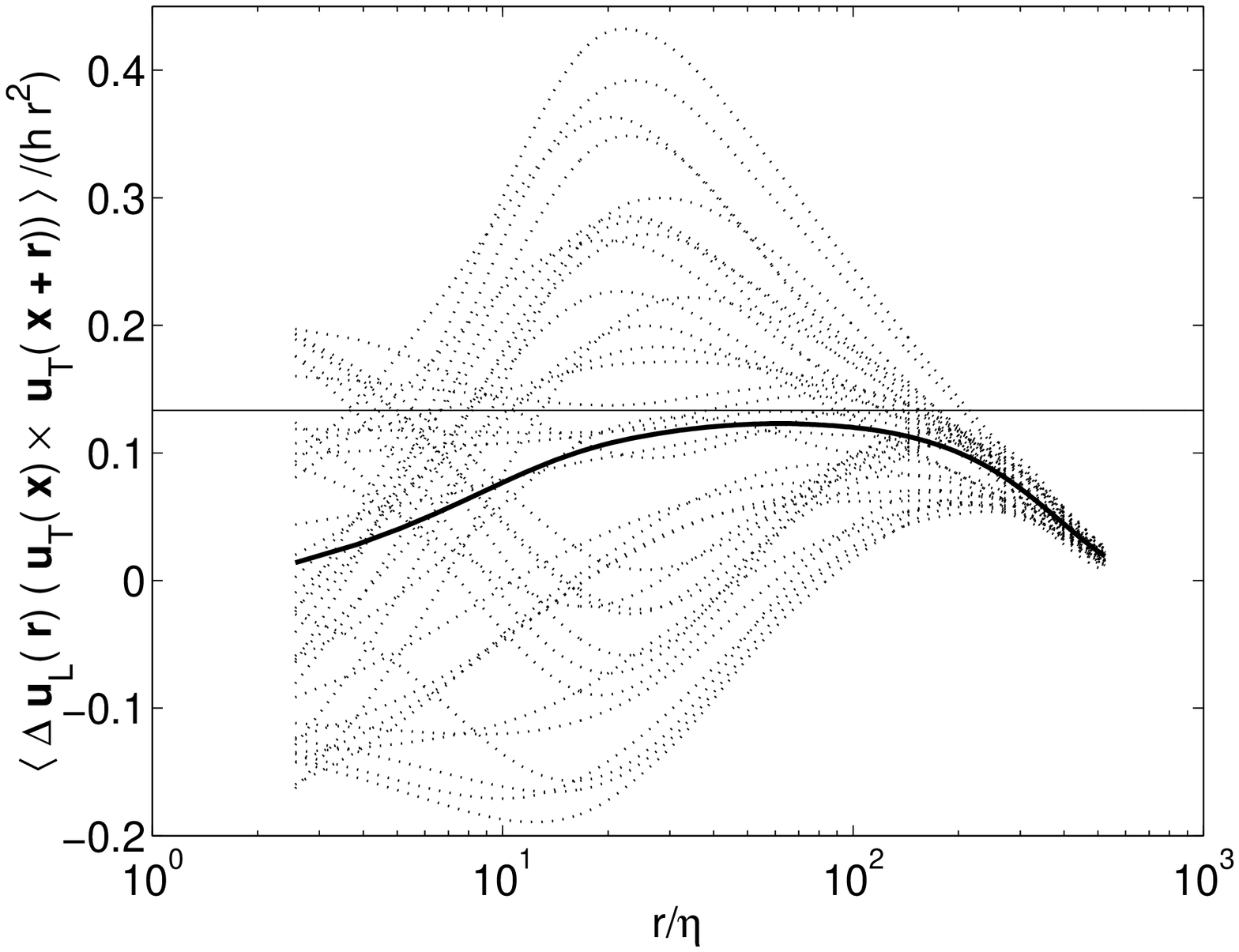}}
\caption{Instantaneous and time-averaged 2/15 law calculations. The
  2/15 value is indicated by the horizontal line in both plots.  The
  inertial range is roughly
  estimated to be $30 < r/\eta < 150$. Note that the vertical scale in the
  two plots is not the same; the time-averaged quantities on the right
  have reduced spread compared to the instantaneous
  quantities on the left.}
\label{k215}
\end{figure}
We first define the compensated quantity
\begin{equation}
\widetilde H_{LTT}({\bf r}) = H_{LTT}({\bf r})/(h~r^2). 
\end{equation}
In Fig.~\ref{k215_frame} we show $\widetilde H_{LTT}({\bf r})$
calculated from a single frame arbitrarily chosen after the flow
achieved statistically steady state. Each dotted line is $\widetilde
H_{LTT}({\bf r})$ in one of 73 different directions in the flow, as a
function of scale size $r$. The directions are fairly uniformly
distributed over the sphere (see \cite{TayKurEyi03} for how these
directions were chosen). None of the curves show a tendency towards
the theoretically predicted $2/15 = 0.1$\.3 value for an extended
range of scales. Among the calculations shown are those for the three
coordinate directions which are the most often reported in statistical
turbulence studies. For any given $r$ the different directions yield
vastly different values.  Exceptional are the very largest (forced)
scales where the different directions appear to converge. This already
signals something different than the usual expectation that
anisotropy, if any, should come from, and dominate in, the large
scales. The anisotropy persists strongly into the smallest resolved
scales, as seen in the large spread of values among the different
directions at $r/\eta \approx 2$, where we might expect viscous
effects are important. Indeed it appears that it would be fortuitous
for the statistics in an arbitrary direction to yield the correct
theoretical prediction for isotropic flow.

Next we extract the isotropic component of these statistics using the
angle-averaging technique of \cite{TayKurEyi03}. The resulting
angle-independent contribution is the thick solid curve in
Fig.~\ref{k215_frame}. Its peak value is $\approx 0.124$, within 7\%
of the 2/15 value. While individual directions are both
parity-breaking as well as anisotropic, the angle-averaged value
recovers the isotropic component of the parity-breaking features
(recall the analogy to the box of screws in Section 1).  This is a
remarkable result out of a single snapshot; there is no $a$ $priori$
reason to expect that angle-averaging an arbitrarily chosen, highly
anisotropic snapshot, will yield consistency with the 2/15-law which
was derived for isotropic flow. We believe that this result is strong
motivation for the existence of a local 2/15 law analogous to the
local 4/5-law of \cite{Eyink03},
\begin{eqnarray}
\langle \Delta u_L({\bf r})(u_T({\bf x}+{\bf r})\times u_T({\bf
  x}))\rangle_{(\Omega,B)} &=& 
\lim_{\delta\rightarrow 0}\frac{1}{\delta}\int^{t+\delta}_t d\tau
\int \frac{d\Omega}{4\pi}\int_B\frac{d{\bf x}}{R^3}\nonumber\\
&\times&[\Delta u_L({\bf r};{\bf x},\tau)][(u_T({\bf x}+{\bf r}) \times u_T({\bf x}))]\nonumber\\
&=&\frac{2}{15}h_B r^2.\label{215_local}
\end{eqnarray}
where $\Omega$, $B$, $R$ and $t$ have the same meanings as for
Eq.~(\ref{eyink45}); $h_B$ denotes the locally (in space and time)
averaged helicity dissipation rate. We emphasize that there is as yet
no proof for Eq.~(\ref{215_local}); we have merely written down a
conjecture by analogy to Eq.~(\ref{eyink45}), motivated by the
calculations for a single snapshot of the flow.

To check if the anisotropy observed in a single frame persists over
time, we averaged $\widetilde H_{LTT}({\bf r})$ in each of the 73
different directions over 9 large-eddy turnover times (45 frames). We
performed the same time-average for the angle-average. The result is
shown in Fig.~\ref{k215_timeave}. The spread in the inertial-range
decreased by about a factor of 2 while the spread in the smallest
scales decreased by a factor of about 6 relative to the single-frame
statistics of Fig.~\ref{k215_frame}.
\begin{figure}
  \centering \includegraphics[scale = 0.5]
  {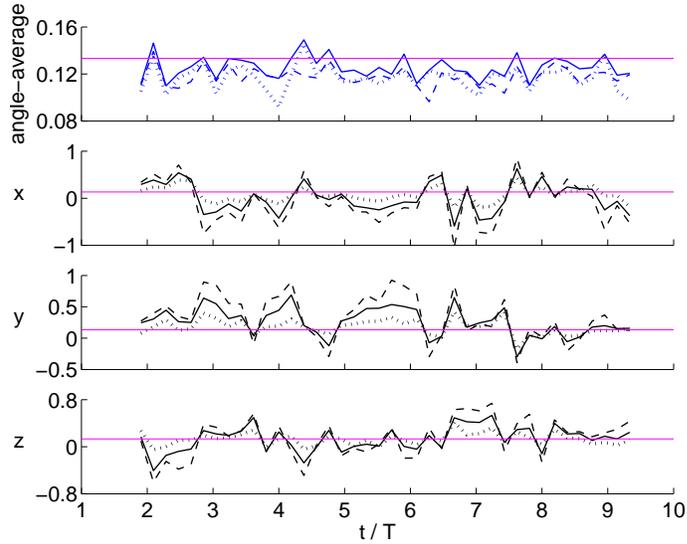}
\caption{$\widetilde H_{LTT}({\bf r})$ for various inertial range values
  of $r/\eta$ as a function of number of eddy turnover times. $r/\eta
  = 30$ (dashed line), $r/\eta = 65$ (solid line), $r/\eta = 120$
  (dotted line). From top to bottom: Angle-averaged, x-direction,
  y-direction, z-direction. Note that the vertical scales in the four
  panels are not the same; the bottom three panels corresponding to
  the coordinate directions have a much greater spread of values than
  the top panel for the angle-average. The mean and standard
  deviations for the $r/\eta = 65$ (middle of the inertial range) in
  each case are given in Table~\ref{peak_table}. \label{k215_time}}
\end{figure}
\begin{table}
\centering
\begin{tabular}{ccc}
Inertial range  & 2/15-law & 4/5-law\\[3.5pt]
Theory & 0.13\.3 & 0.8\\
Angle-avg & 0.126 $\pm$ 0.009 & 0.75 $\pm$ 0.03\\
x & 0.02 $\pm$ 0.31 & 0.78 $\pm$ 0.14\\
y & 0.26 $\pm$ 0.23 & 0.75 $\pm$ 0.13\\
z & 0.14 $\pm$ 0.21 & 0.76 $\pm$ 0.11\\
\end{tabular}
\caption{Mean and standard deviation of the compensated third order statistics in the middle of the inertial range.\label{peak_table}}    
\end{table}
Inspite of this, the residual variance is significant as we
demonstrate in Fig.~\ref{k215_time} and as compared below to the same
analysis done for the 4/5-law. We plot a time-trace of the
angle-averaged $\widetilde H_{LTT}(r)$ in the top panel of
Fig.~\ref{k215_time} for $r/\eta = 30$ (lower end of the inertial
range), $r/\eta = 65$ (middle of the inertial range) and $r/\eta =
120$ (higher end of the inertial range). The angle-averaged value in
the middle of the inertial range ($r/\eta = 65$) is
0.126$\pm$0.009, within error of the predicted value of 2/15 $=$
0.1\.3. The value ranges from $0.119$ and $0.126$ across the inertial
range with variances of $8-9\%$. This puts the mean angle-averaged
value within 1.5 standard deviations of the theoretical value of 2/15
across the inertial range. Since most prior numerical simulations
investigations have studied two-point statistics in the coordinate
directions only, we present in the bottom three panels of
Fig.~\ref{k215_time}, the values of $\widetilde H_{LTT}( {\bf r})$
again at various values of $r/\eta$ for ${\bf {\hat r}}$ in the $x$-,
$y$- and $z$-directions respectively as a function of time.
Table~\ref{peak_table} (column 2) shows the mean and standard
deviation for each of the four time-trace plots of
Fig.~\ref{k215_time} in the middle of the inertial range at $r/\eta =
65$. The first thing to notice is that none of the coordinate
directions average to $2/15$ over long times. This behavior
demonstrates that these statistics are highly anisotropic and remain
so over long times. Secondly, the mean values in the coordinate
directions are poorly defined and practically meaningless in the sense
of having extremely large standard deviations. In turbulence
phenomenology, such large jumps in values from their mean is the
signature of intermittency--the presence of strong, anomalous events.
We conclude that the helicity flux in a particular direction is highly
intermittent in the inertial range (see \cite{CCEH03} for a different
approach to this issue).

\begin{figure}
  \centering{\label{k215d_frame}
    \includegraphics[scale=0.35]{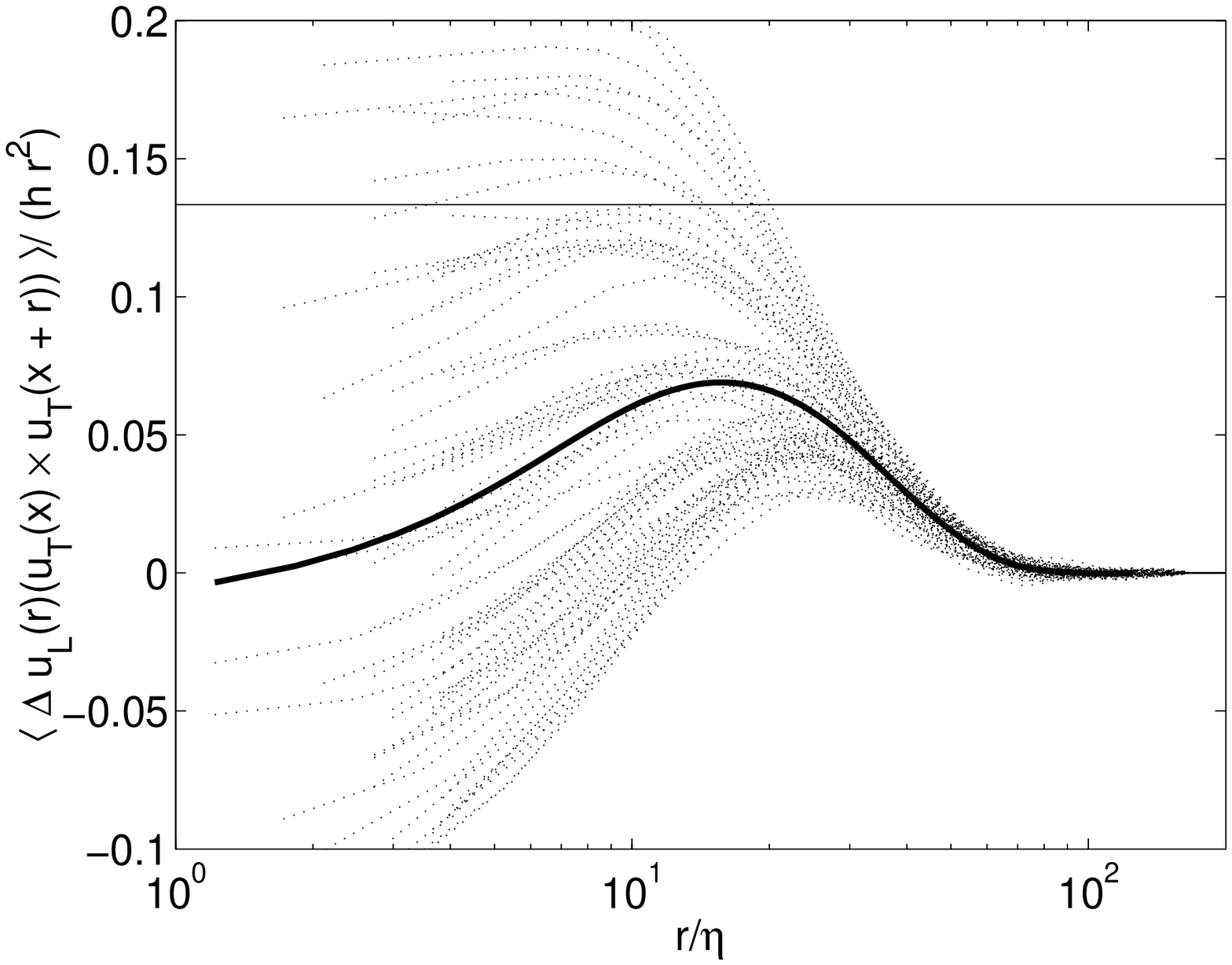}}
  {\label{k215d_timeave}
    \includegraphics[scale=0.35]{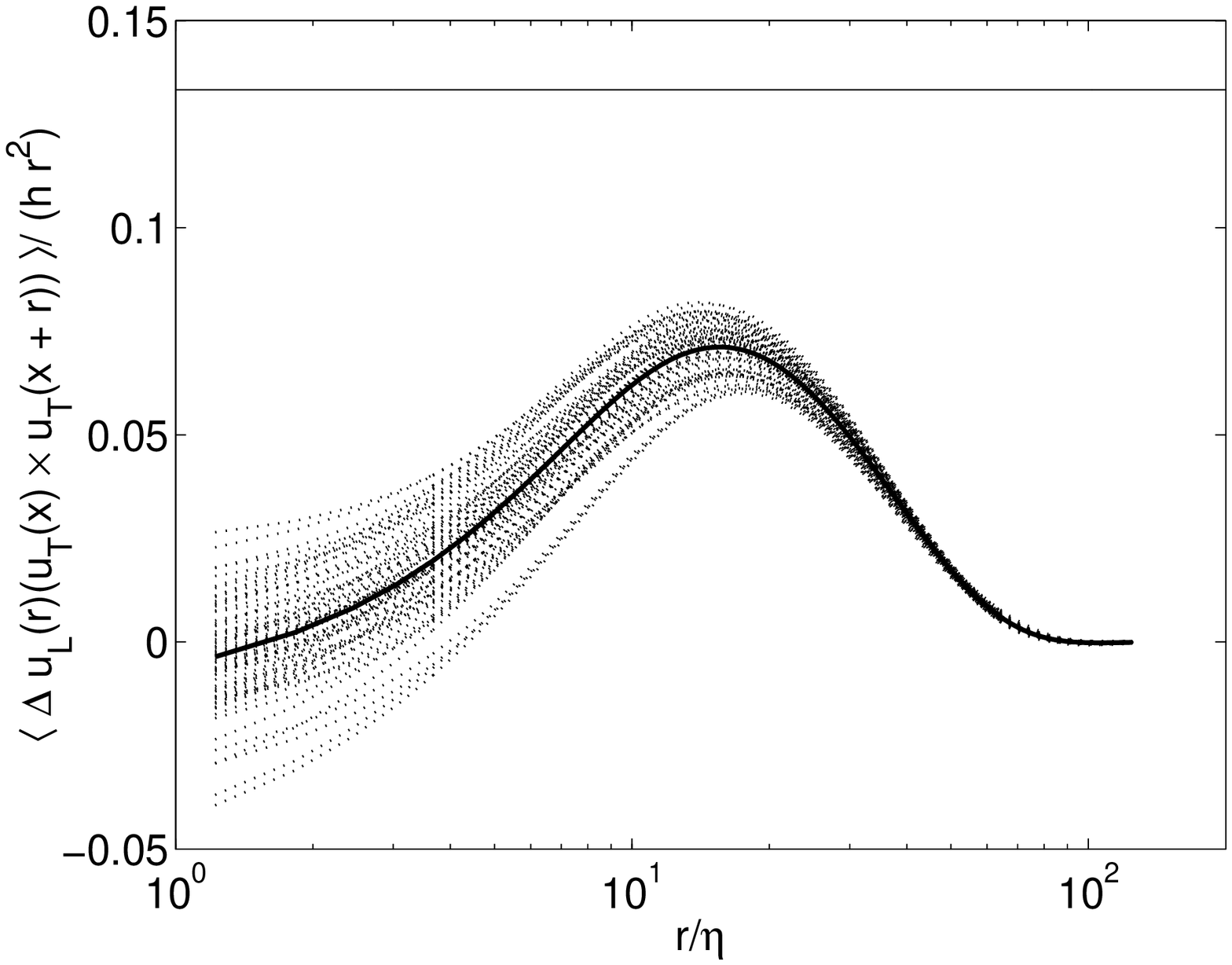}}
  \caption{Instantaneous (left-panel) and ensemble-averaged
    (right-panel), calculations of $\widetilde H_{LTT}({\bf r})$ for
    the test decaying case at $256^3$ ($R_\lambda \sim 50$). The
    initial energy and helicity spectra were isotropic with helicity
    in all the modes. The snapshot is from when the flow had maximally
    developed and the ensemble average is over 30 such realizations
    with random phases for each realization. The 2/15 value is
    indicated by the horizontal line in both plots. Note the
    qualitative similarity to the forced case of Fig.~\ref{k215}
    especially in the small scales ($r < 20$).  The 2/15 value is not
    attained, presumably due to the low Reynolds number. }
\label{k215d}
\end{figure}

We return briefly to one of the subtleties of the 2/15-law, namely
that it was derived for decaying flows. Our initial tests of the
2/15-law in ensemble averages over decaying flows with prescribed
initial isotropic helicity and energy spectra (see \cite{PolShtil89}
for the method) showed qualitatively the same results. That is, the
large anisotropy among the different directions and their
intermittency is similar to the forced case reported here and their
angle-average has the same qualitative behavior indicating some
constant flux in the middle of the range and smoothly approaching
zero as $r \rightarrow 0$ (see Fig.~\ref{k215d}).
However the Taylor Reynolds numbers achieved were too low ({\cal
  O}(50)) to see the 2/15 value which is our primary interest in this
investigation. Our computational resources restrict us to forced
turbulence when investigating high-Reynolds number effects. Given the
qualitative similarities in the data, we anticipate that a careful
analysis for the forced case would give the same result as for the
decaying case.

\subsection{Comparative analysis of the 4/5-law}
We compare these results with the analogous ones for the 4/5-law for
the same flow. As before, we define the compensated third-order
longitudinal structure function
\begin{equation}
\widetilde S_{L,3}({\bf r}) = \langle (u_L({\bf x}+{\bf r}) -
u_L({\bf x}))^3 \rangle /(\varepsilon r).
\end{equation}
\begin{figure}
\centering
\subfigure[$\widetilde S_{L,3}({\bf r})$ in 73 different directions of the flow
(dotted lines) from an arbitrarily chosen frame after the flow has
achieved statistical equilibrium. The thick solid line is the mean
over all the directions.]{
\label{k45frame}\includegraphics[scale=0.35]{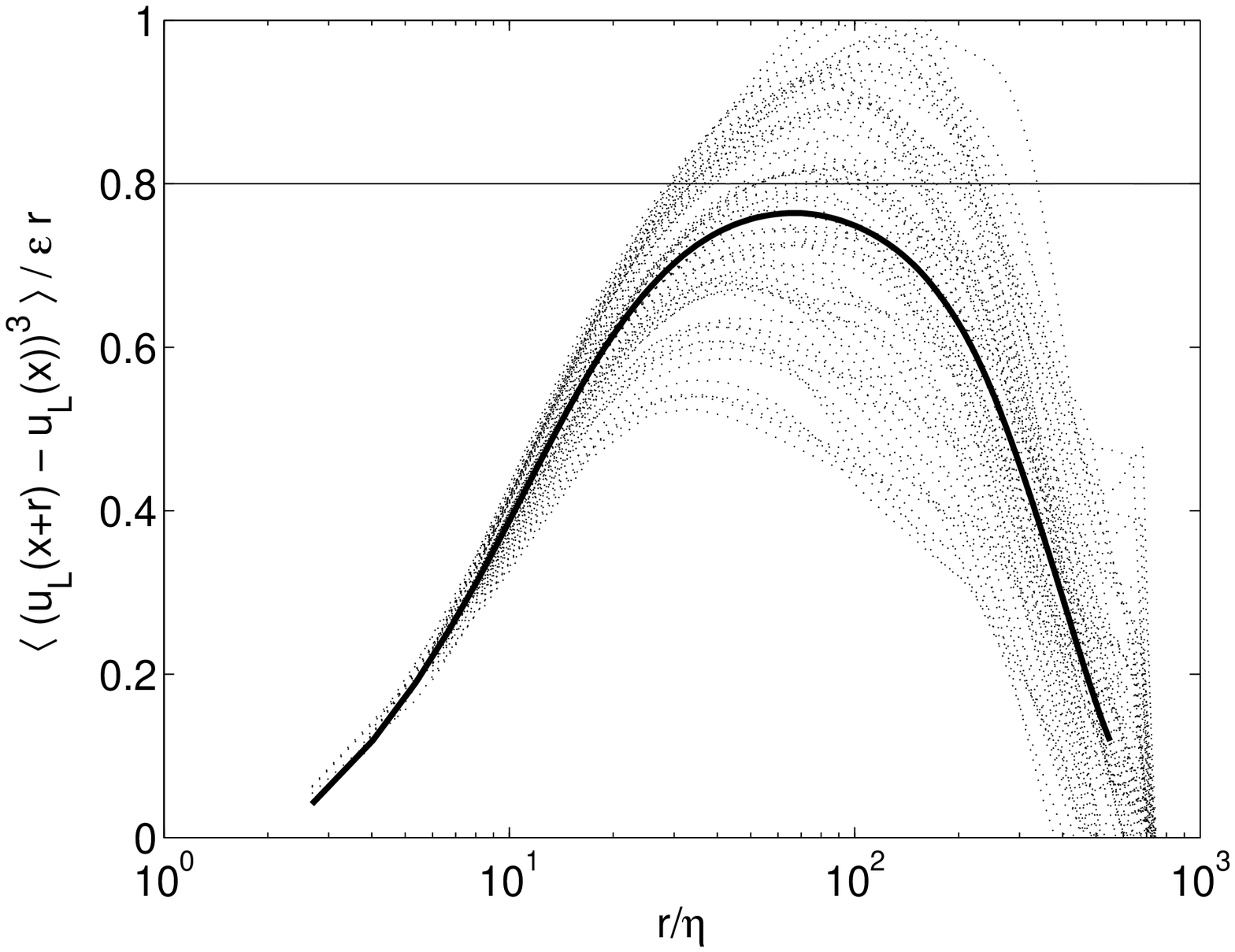}
}\hspace{-0.1in}
\subfigure[The time-average of $\widetilde S_{L,3}({\bf r})$ in each of
  the 73 different directions (dotted lines) and of 
  the angle-average (thick solid line).]{
\label{k45avg}\includegraphics[scale=0.35]{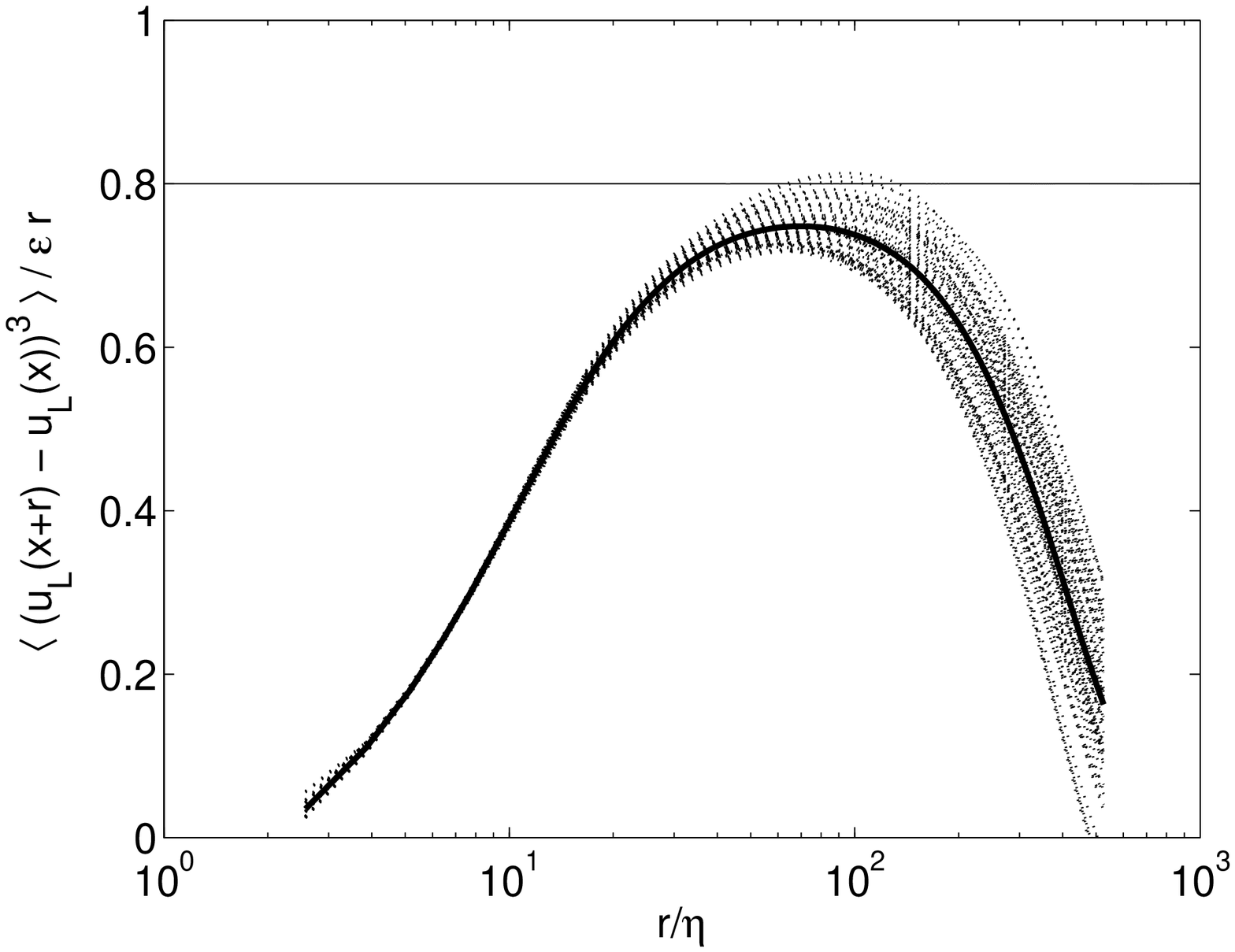}
}
\caption{Instantaneous and time-averaged calculations of the
  4/5-law. The 4/5 value is indicated by the horizontal line in both plots.}
\label{k4/5}
\end{figure}
Figure \ref{k45frame} shows a single frame calculation of $\widetilde
S_{L,3}({\bf r})$ for 73 different directions as a function of $r$
(dotted lines). We performed the angle-averaging exactly as in
\cite{TayKurEyi03} to recover the isotropic mean (thick solid line).
Our first observation is that the 4/5-law is recovered in this helical
flow to as good a degree as in the simulation with zero mean helicity
of \cite{TayKurEyi03}. This demonstrates that the reflection-symmetry
assumption of Kolmogorov need not hold in order to see this result.
This is understood by the fact that the lowest-order (unclosed)
dynamical equations for the symmetric second-order correlation
functions (\cite{KH38}) from which the 4/5-law was derived by
\cite{K41b}, decouple from their antisymmetric counterpart
(\cite{Betchov61}) from which the 2/15-law was derived.  The
third-order longitudinal structure functions of the K41 law are
reflection-symmetric by definition, and therefore cannot directly
probe the helical, parity-breaking properties of the flow; while the
third-order correlation function $H_{LTT}$, cannot probe the
reflection-symmetric properties of the flow.  The 4/5- and 2/15-laws
in fact coexist in turbulent flows with helicity. This possibility was
first hinted at by \cite{Betchov61}, who noted that in the equations
of motion of statistical moments, the fourth-order correlation function
dynamics are the lowest order at which coupling of the symmetric
(energy-dependent) and antisymmetric (helicity-dependent) quantities
can occur.

In Fig.~\ref{k45avg} the
time-averaged compensated third-order longitudinal structure function
for all the directions converge rather well relative to the single
frame in Fig.~\ref{k45frame} in the inertial range. 
There is still significant spread of
values among the different directions in the inertial range but it is
far less than in the time-averaged 2/15-law calculation in Fig.
\ref{k215_timeave}.  To make a more quantitative comparison, we
present the mid-inertial-range values of the angle-averaged, and the
$x$-, $y-$ and $z-$direction calculations in Table~\ref{peak_table},
column 3. The time mean for the angle-average is well-defined at
$0.75\pm0.03$, a small standard deviation of 4$\%$. The means in the
coordinate directions range from $0.75$ to $0.78$, not intolerably far
from the $0.8$ value expected from theory, but with significant
standard deviation in time of the order of 20$\%$. Still, the behavior
is very different from the 2/15-law statistics
(Table~\ref{peak_table}, column 2), where not only was the 2/15 value
not achieved in an arbitrary direction, but the variability in time
was huge, 100\% or more. We are lead to conclude that in the inertial
range, both instantaneously and over long times, the helicity flux (as
described by the 2/15-law) is far more anisotropic and intermittent
than the energy flux (as described by the 4/5-law) for the same
statistically steady flow.

\subsection{The viscous range}\label{vis-range}
Anisotropy of $\widetilde H_{LTT}$ persists into the
smallest resolved scales as demonstrated by the large variance among
the directions in the range $r/\eta <10$ in Figs.~\ref{k215_frame} and
~\ref{k215_timeave}. By contrast, the angular-dependence of
$\widetilde S_{L,3}({\bf r})$ becomes very small in the same range in
a snapshot (Fig.~\ref{k45frame}) and even more so on average over time
(Fig.~\ref{k45avg}). In these scales the 2/15- and 4/5-laws no longer
hold as viscous effects become important; the quantities $\widetilde
H_{LTT}$ and $\widetilde S_{L,3}$ no longer correspond strictly to the
helicity and energy fluxes respectively. The viscous terms for the
symmetric quantities, interpreted as energy dissipation at scales $r
\approx \eta$, are strictly a sink for energy, pulling energy out of
the flow. As is well known, the viscous terms for the antisymmetric
quantities, correspondingly the helicity-dissipation at scales $r
\approx\eta$, may be positive (producing helicity) or negative
(removing helicity). Nevertheless, if the small scale statistics
$H_{LTT}$ are to be isotropic, the different directions might be
expected to converge in the very small scales. In
Table~\ref{min_table}, column 2, we show time-mean and standard deviation
of the angle-average and the $x$-, $y$- and $z$-direction calculations
of $H_{LTT}({\bf r})$ at $r/\eta \approx 2$.  The time-mean angle-averaged
value is about 0.014 $\pm$ 0.004, a standard deviation of about
$30\%$. As in the inertial range, the time-mean in a particular
direction does not agree with the angle-averaged value and the standard
deviations are enormous. We have
shown the corresponding numbers for the 4/5-law for comparison
(Table~\ref{min_table}, column 3); the means in a particular direction
agree better with the angle-averaged mean, and the variances are
around 5\%, indicating recovery of isotropy in the small-scales and
relatively weaker intermittency than for the 2/15-law statistics.  We
here introduce a note of caution about the results in the viscous
range as our simulation is only resolved upto $r/\eta \approx 2$
($k_{\rm max}/\eta \approx 1.1$).
While the inertial range is amply resolved, the viscous range might
display some residual effects of being under-resolved. Nevertheless,
to the extent that in the $same$ flow, the energy-dependent statistics
recover isotropy rather quickly in the viscous scales, it
seems worthwhile to note that the helicity-dependent statistics
remain dramatically and persistently anisotropic in the viscous scales
over the long duration of our simulation.

\begin{table}
\centering
\begin{tabular}{ccc}
Viscous range  & 2/15-law & 4/5-law\\[3.5pt]
Angle-avg & 0.014 $\pm$ 0.004 & 0.035 $\pm$ 0.006\\
x & -0.05 $\pm$ 0.65 & 0.057 $\pm$ 0.004 \\
y & 0.02 $\pm$ 0.57 & 0.055 $\pm$ 0.003\\
z & 0.17 $\pm$ 0.55 & 0.057 $\pm$ 0.003\\
\end{tabular}
\caption{Mean and standard deviation of the compensated third order
statistics for the smallest resolved scale. \label{min_table}}
\end{table}

\section{Conclusions}
This analysis shows that the statistically steady
state (forced turbulence) as well as $local$ versions of the 2/15-law
analogous to the forced (\cite{Frisch95} and local 4/5-law
(\cite{Eyink03}), might hold true; we hope our empirical
results motivate a theoretical effort towards the proofs. The
helicity flux is significantly more anisotropic and intermittent than
the energy flux. This suggests that the viscous generation and
dissipation of helicity in the small scales is highly anisotropic as
well. This might be related to the strong helical events seen in the
transverse alignment of vortices in the work of \cite{HolmKerr02}.
There is an underlying isotropic component of the flow which is
extracted by the angle-averaging procedure of \cite{TayKurEyi03}. It
is not surprising that angle-averaging recovers the
orientation-independent component of the field. However it is
remarkable that this spherically averaged value tends to the predicted
4/5 and 2/15 values for $\widetilde S_{L,3}$ and $\widetilde H_{LTT}$
respectively. This suggests that the 'local isotropy' requirement of
K41 may be relaxed in favor of a hypothesis that the flow statistics
have a universal underlying isotropic component.

We conclude with two remarks which were not explicitly mentioned in
the body of this paper. The issues of anisotropy and intermittency of
the small-scales of the flow are intimately connected with the
particular kind of statistics measured. We have shown that in the same
flow, certain statistics which depend on energy flux recover isotropy
in the small scales, while others which depend on helicity do not. It
is therefore more sensible to speak of isotropy (or lack of isotropy)
of the $statistics$ of the flow rather than of the flow itself.  A
second relevant remark is that our numerical data and analysis give
some indication as to what might be expected when measuring
$H_{LTT}({\bf r})$ in high-Reynolds number experimental flows. In many
such experiments, data is acquired at a few points over long times,
and the statistics are obtained by applying Taylor's hypothesis to
obtain the spatial correlations in a single-direction (for example,
the streamwise direction in a windtunnel). Assuming there is some
helicity in the flow, it might not be possible to predict the behavior
of $H_{LTT}({\bf r})$ for a particular direction ${\bf \hat r}$ (see
\cite{KhoShaTsi01}). In this respect, the full-field information and
angle-averaging technique appear to be $fundamental$ to recovering the
2/15 isotropic prediction. An experimental effort such as the
three-dimensional velocity field imaging of \cite{TaoKatMen02} might
be needed to see the $2/15$-law experimentally.  This is very
different from measurement of the 4/5-law for energy, where, given
Reynolds number high enough, the statistics in any direction are
observed to recover isotropy in the small scales.

\acknowledgements
We are grateful for useful discussions with D.D.~Holm and G.L. Eyink.
\bibliography{two15law_jfm}
\bibliographystyle{jfm}
\end{document}